\documentclass[a4paper,11pt]{article}
\pdfoutput=1 

\usepackage{multirow}
\usepackage{jinstpub} 

\title{Detector and Physics Performance at a Muon Collider}

\author[1] {Nazar Bartosik,}
\author[2] {Alessandro Bertolin,}
\author[6] {Laura Buonincontri,}
\author[3] {Massimo Casarsa,}
\author[4] {Francesco Collamati,}
\author[5] {Alfredo Ferrari,}
\author[8] {Anna Ferrari,}
\author[2] {Alessio Gianelle,}
\author[6] {Donatella Lucchesi,}
\author[9] {Nikolai Mokhov,}
\author[10]{Mark Palmer,}
\author[1] {Nadia Pastrone,}
\author[7] {Paola Sala,}
\author[2] {Lorenzo Sestini,}
\author[9] {Sergei Striganov}

 \affiliation[1]{INFN Sezione di Torino, Torino, Italy}
  \affiliation[2]{INFN Sezione di Padova , Padova, Italy}
  \affiliation[3]{INFN Sezione di Trieste , Trieste, Italy}
  \affiliation[4]{INFN Sezione di Roma , Roma, Italy}
  \affiliation[5]{CERN, Geneva, Switzerland}
  \affiliation[6]{University of Padova and INFN Sezione di Padova , Padova, Italy}
  \affiliation[7]{INFN Sezione di Milano, Milano, Italy}
  \affiliation[8]{HZDR, Dresden, Germany}
  \affiliation[9]{Fermilab, Batavia, Illinois, United States}
  \affiliation[10]{Brookhaven National Laboratory, Upton, New York, United States}

\emailAdd{donatella.lucchesi@pd.infn.it, lorenzo.sestini@pd.infn.it, massimo.casarsa@ts.infn.it, nazar.bartosik@to.infn.it}
\abstract{A muon collider represents the ideal machine to reach very high center-of-mass energies and luminosities by colliding elementary particles. This is the result of the low level of beamstrahlung and synchrotron radiation compared to linear or circular electron-positron colliders. In contrast with other lepton machines, the design of a detector for a multi-TeV muon collider requires the knowledge of the interaction region due to the presence of a large amount of background induced by muon beam decays. The physics reaches can be properly evaluated only when the detector performance is determined. In this work, the background generated by muon beams of $750$ GeV is characterized and the performance of the tracking system and the calorimeter detector are illustrated. Solutions to minimize the effect of the beam-induced background are discussed and applied to obtain track and jet reconstruction performance. The $\mu^+\mu^-\to H\nu\bar{\nu}\to b\bar b \nu\bar{\nu}$ process is fully simulated and reconstructed to demonstrate that physics measurements are possible in this harsh environment. The precision on Higgs boson coupling to $b\bar b$ is evaluated for $\sqrt{s}=1.5$, 3, and 10 TeV and compared to other proposed machines.}
\arxivnumber{1234.56789} 
\begin{document}
\maketitle
\flushbottom
\section{Introduction}
\label{sec:intro}
In a multi-TeV muon collider, the effects of the background induced by the muon beam decays need to be evaluated in detail at the detector level to be able to estimate the physics reach.
In fact, the muon decay products can contaminate the Interaction Region (IR) from a distance that varies with the beam energy and the collider optics and its superconducting magnets, with appropriate protective elements, need to be designed and included in the simulations~\cite{M1} to have an accurate background description. Previous studies~\cite{M2,mokhov15,mokhovHF} have found that two cone-shaped tungsten shields can be used to protect the IR and the detector from the extremely high flux of beam background particles. As a consequence, the exact design of the machine-detector interface (MDI), which includes these shields, is needed to evaluate the distribution of the induced background at any part of the detector. 
The beam-induced background generated by the MAP collaboration~\cite{mapc} with the IR and the MDI optimized for 1.5-TeV center-of-mass energy~\cite{mokhov15} are used to study the effect on the tracking system and on the calorimeter detector. 
The software framework used for the propagation of the beam-induced background through the detector is inherited from the MAP collaboration and has been further developed to obtain the final results.
The production and decay of the Higgs boson $\mu^+\mu^-\to H\nu\bar{\nu}\to b\bar b \nu\bar{\nu}$ at $\sqrt{s}=1.5$~TeV is studied to determine the reconstruction and identification efficiencies, that are used to evaluate the sensitivity to the $b\bar b$ coupling. Samples of $\mu^+\mu^-\to H\nu\bar{\nu}\to b\bar b \nu\bar{\nu}$ at particle level are generated also at $\sqrt{s}=3, 10$~TeV and a conservative calculation of the sensitivities at these energies is presented.
\section{Beam-induced background characterization}
\label{sec:bckstudy}
\begin{figure}[htbp]
  \centering
    \includegraphics[width=0.6\textwidth]{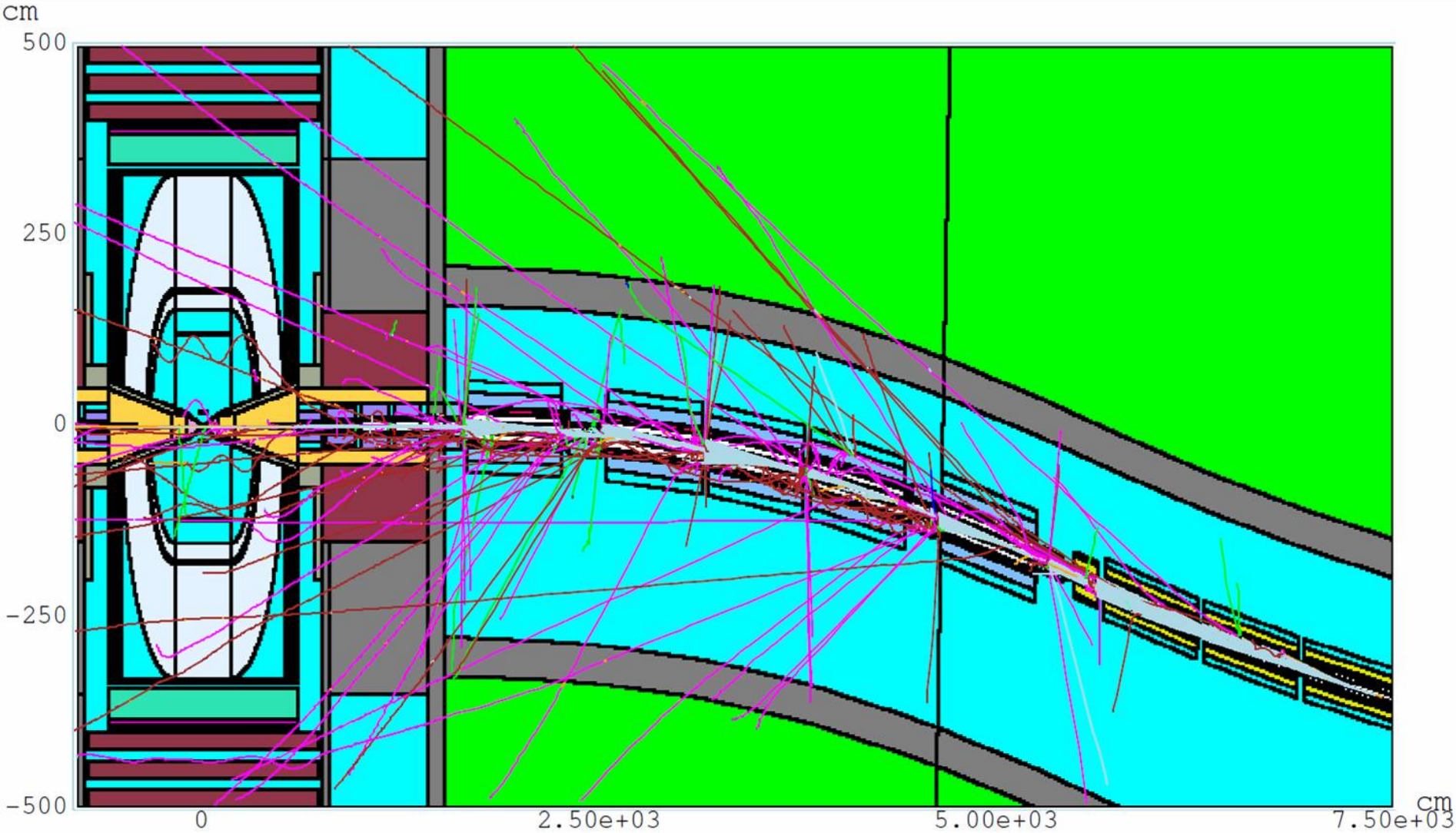}
    \caption{Illustration of the model of the machine and machine-detector interface built for the MARS15 simulation. The shielding nozzles, described in the text, are represented in yellow inside the detector. This figure has been reproduced from Ref.~\cite{vito}.
    \label{fig:mars_model}}
\end{figure}

The simulation of the beam-induced background has been performed with the MARS15 software~\cite{MARS15} and studied for machines with a center of mass energy of $\sqrt{s}=1.5$~TeV and $\sqrt{s}=125$~GeV, as discussed in detail in Refs.~\cite{mokhov15,mokhovHF} and in Ref.~\cite{mc-pre}.
The background particles reaching the detector are mainly produced by the interactions of the decay products of the muon beams with the machine elements. Their type, flux, and characteristics strongly depend on the machine lattice and the interaction point configuration, which in turn depend on the collision energy. 
The background particles may be produced tens of meters upstream the interaction point, as can be seen in Figure~\ref{fig:mars_model}. Therefore, a detailed layout of the machine and the machine-detector interface must be included in the simulation.
In particular, the opening angle of two shielding cones (``nozzles''), introduced to mitigate the effects of the beam-induced background inside the detector, must be optimized for a specific beam energy and it will affect the detector acceptance.

\begin{figure}[htbp]
  \centering
    \includegraphics[width=0.32\textwidth]{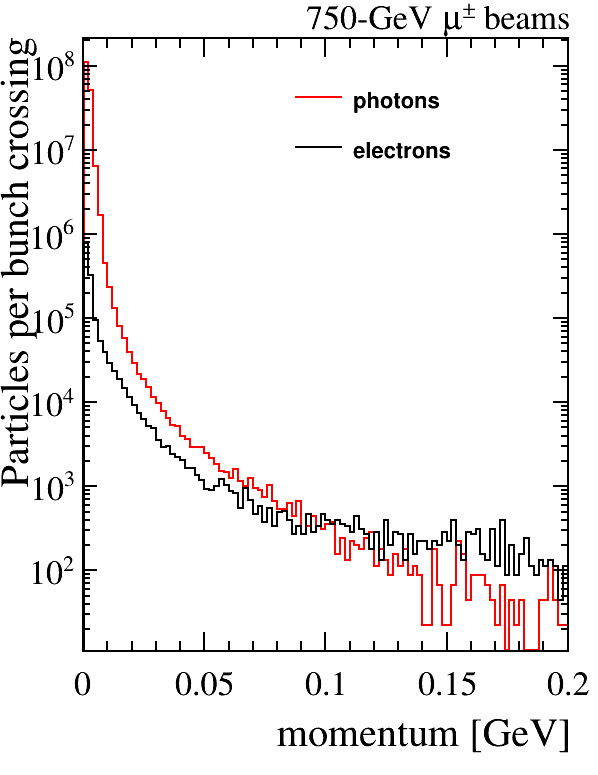}
    \includegraphics[width=0.32\textwidth]{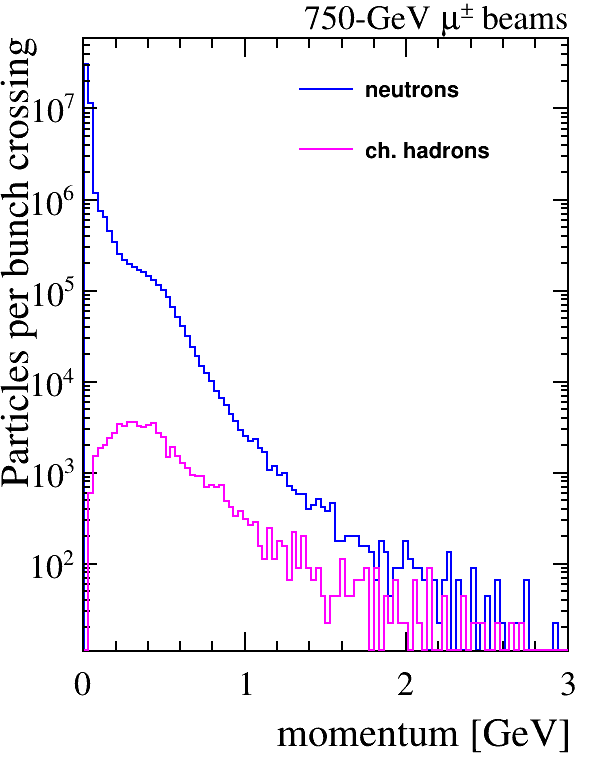}
    \includegraphics[width=0.32\textwidth]{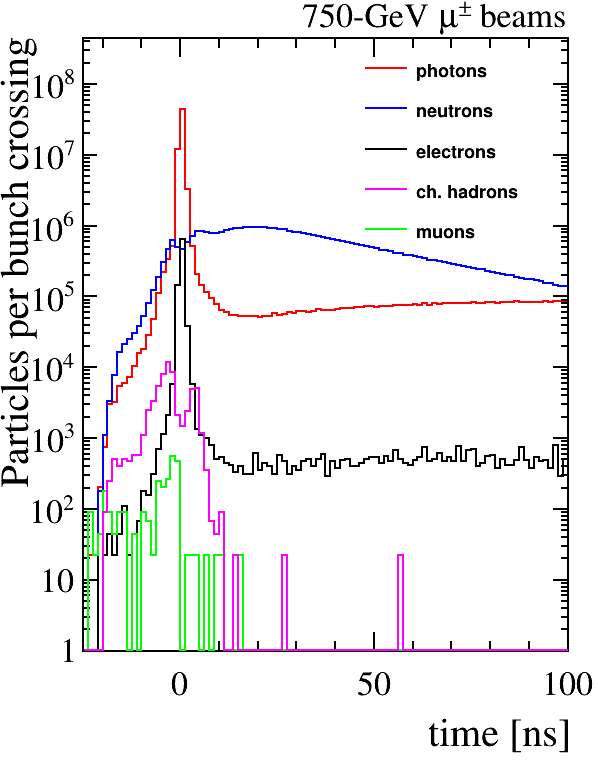}
    \caption{Characteristics of the beam-induced background particles at the detector entry point: the momentum spectra for photons and electrons and for neutrons and charged hadrons are shown in the left and central panels, respectively; the time of arrival with respect to the beam crossing time is shown on the right.
    \label{fig:bkg_char}}
\end{figure}

Nevertheless, the flux of particles surviving the shielding is very high. Their main properties are a relatively low momentum and an arrival time in each sub-detector asynchronous with respect to the beam crossing. Figure~\ref{fig:bkg_char} shows the momentum spectra of the electromagnetic (left) and hadronic (central) components of the beam-induced background for a muon beam of 750~GeV. The first one is relatively soft ($\langle p_{\mathrm{ph.}} \rangle = 1.7$ MeV and $\langle p_{\mathrm{el.}} \rangle = 6.4$ MeV), whereas the second one has an average momentum of about half a GeV ($\langle p_{\mathrm{n}} \rangle = 477$ MeV and $\langle p_{\mathrm{ch. had.}} \rangle = 481$ MeV). The time of arrival of the particles at the detector entry point with respect to the bunch crossing time for the different background components is shown on the right of Figure~\ref{fig:bkg_char}. The peaks that are evident around zero are primarily due to the leakage of photons and electrons around the interaction point, where the shielding is minimal.
\section{Detector performance}
\label{sec:dete}
\begin{figure}[htbp]
\centering
    \includegraphics[width=0.6\textwidth]{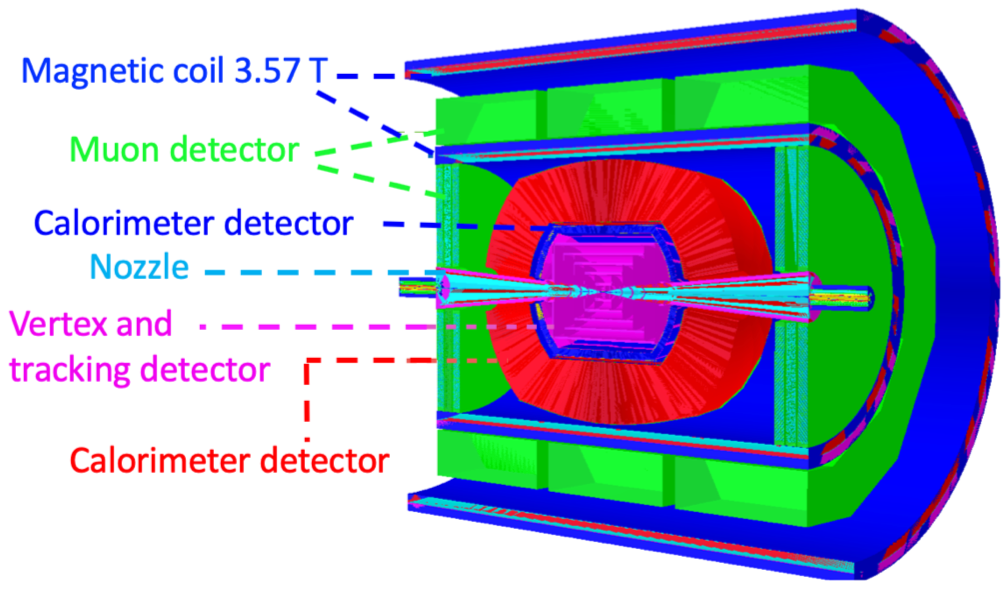}
    \caption{Schematic view of the detector, with each component identified by the label}.
    \label{fig:dete}
\end{figure}

The detector model and software framework used for the studies presented in this paper can be found also in Ref.~\cite{mc-pre} and~\cite{vito}. Figure~\ref{fig:dete} presents a schematic view of the detector components, as implemented in the ILCRoot framework~\cite{IlcRoot}.
These studies focus on the tracking and calorimeter systems, a full simulation of the muon detector is not currently available. Both the tracker and the calorimeter are immersed in a solenoidal magnetic field of 3.57~T.

The tracking system consists of a vertex detector (VTX), an outer silicon tracker (SiT) and a forward tracker (FTD). The vertex detector, located just outside a 400-$\mu$m thick Beryllium beam pipe of 2.2-cm radius, is 42-cm long with five cylindrical layers at distances from 3 to 12.9~cm in the transverse plane to the beam axis and four disks on each side. Outside the VTX, a 330-cm long silicon tracker is comprised of five cylindrical layers at radial distances between 25 and 126 cm and 14 forward disks. 

The forward tracker consists of three disks on each side of the SiT at $z = \pm 120,\: \pm 155,\: \pm 190$ cm, which cover the very forward regions and are properly shaped to host the nozzles.
The silicon sensors of all the tracking detectors are based on silicon pixel technology. The VTX sensors have a pixel granularity of $20\times 20~\mu$m$^2$ and a thickness of $75~\mu$m in the barrel and $100~\mu$m in the disks. The SiT and the FTD feature pixel sensors of $50\times 50$-$\mu$m$^2$ pitch and 200-$\mu$m thickness. 
The full simulation output is digitized, with signals that include electronic noise as well as threshold and saturation effects. 

\begin{figure}[htbp]
\centering
    \includegraphics[width=0.6\textwidth]{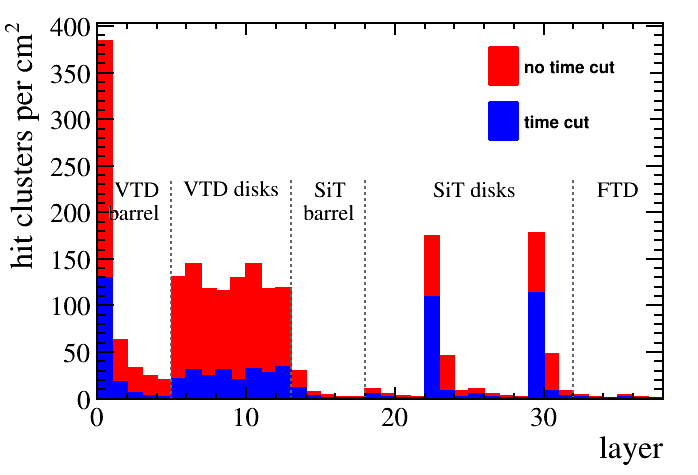}
    \caption{Hit density for each tracker layer before and after the time requirements.}
    \label{fig:Si-occupancy}
\end{figure}

Despite the mitigation of the beam-induced background provided by the shielding nozzles, a large particle flux reaches the detector, causing a very high occupancy in the first layers of the tracking system that impacts the detector performance. 
New generation 4D silicon sensors, which provide both spatial and time information, would allow exploitation of the peculiar time distribution of the background hits and remove a significant fraction of them.
An optimized hit selection time window for each tracking layer was studied in a sample of muons, generated at the IP with a flat $p_T$ spectrum between 0.5 and 100 GeV.
\begin{figure}[htbp]
\centering
    \includegraphics[width=0.48\textwidth]{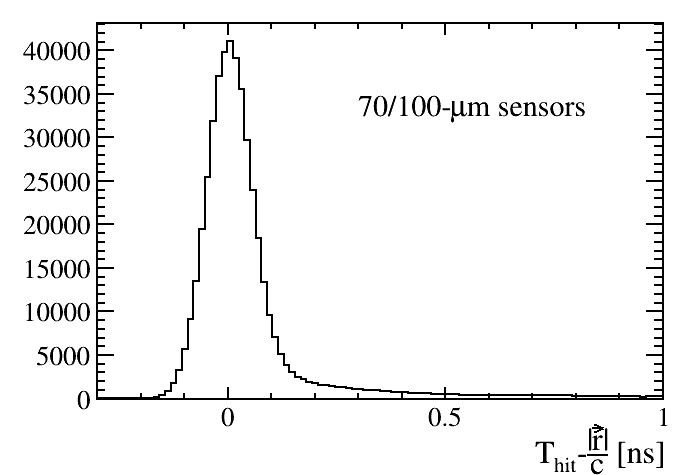}
    \includegraphics[width=0.48\textwidth]{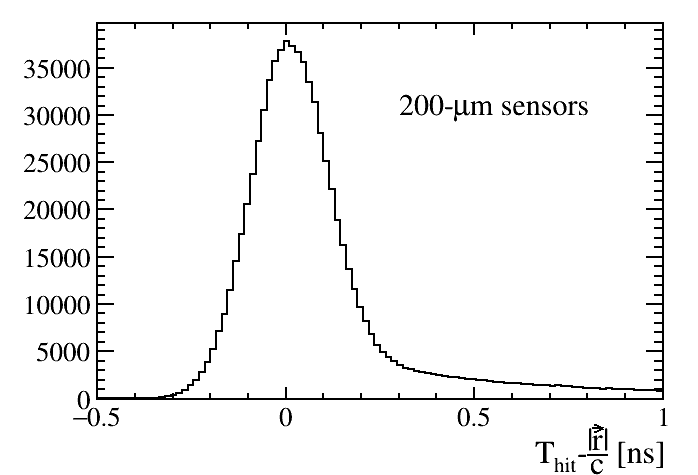}
    \caption{Distributions of the tracker hit times with respect to the bunch crossing for 75~$\mu$m/100~$\mu$m-thick sensors (left) and 200~$\mu$m-thick sensors (right) in a sample of muons  with $0.5 < p_T < 100$ GeV. The tails on the right are due to the low-momentum muons.}
    \label{fig:Si-time}
\end{figure}
The distributions of the arrival times registered by each sensor with respect to the bunch crossing are shown in Figure~\ref{fig:Si-time}, where average time resolutions $\sigma_T$ of $50$~ps and $100$~ps are assumed for the sensors with a thickness of 75~$\mu$m/100~$\mu$m and $200~\mu$m, respectively~\cite{SiTimeRes}.
The contribution to the time resolution due to the time spread of the beam spot is estimated to be $\sim$30~ps.
Figure~\ref{fig:Si-occupancy} shows the reduction in the tracker layers occupancy, when the hit times are required to be consistent with the arrival time of particles produced at the IP within a $\pm3\,\sigma_T$ window.

\begin{figure}[htbp]
\centering
    \includegraphics[width=0.48\textwidth]{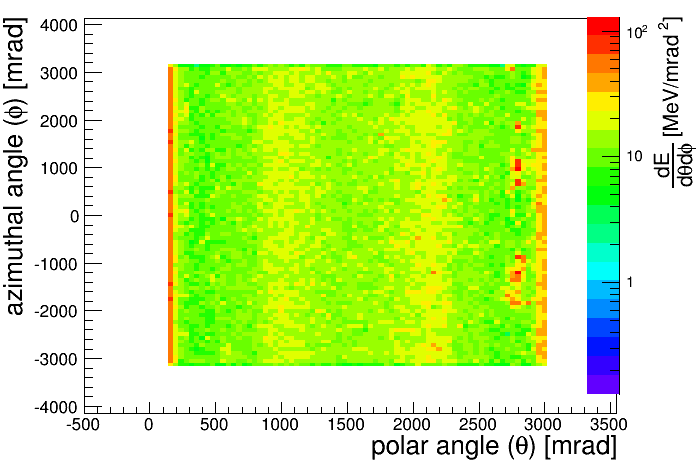}
    \includegraphics[width=0.48\textwidth]{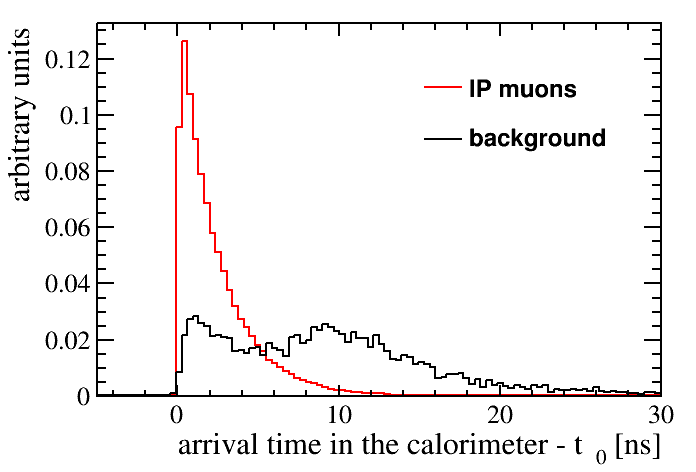}
    \caption{Left: Energy distribution in the calorimeter of the beam-induced background. Right: Distribution of the arrival time at a given calorimetry position of particles generated by beam radiation and of muons coming from primary vertex. The particles arrival time is measured with respect to $t_0$, defined as the arrival time in the given position of a photon produced in the primary vertex at the collision time}.
    \label{fig:calo-occupancy}
\end{figure}

The calorimeter response is fully simulated for a Dual-Readout Integrally Active and Non segmented Option (ADRIANO)~\cite{adriano}. The geometry is fully projective covering polar angles down to $8.4^\circ$. Barrel and end-cap regions are formed by about 23600 towers of $1.4^\circ$ aperture angle of lead glass with scintillating fibers. Cherenkov and scintillation hits are generated separately and digitized independently. The full simulation includes photodetector noise, wavelength-dependent light attenuation and collection efficiency and energy cluster digitization. 
Beam-induced background generates an almost flat distribution of noise in each tower, similar to an underlying event as displayed on the left of Figure~\ref{fig:calo-occupancy}. Part of this noise is removed by the jet clustering algorithm as discussed in~\ref{sec:jet}.
The time of deposited energy in each tower can be exploited further to best suppress the beam-induced background. On the right of Figure~\ref{fig:calo-occupancy} the distribution of arrival time for all the energy cluster is shown in black compared, in red, with the same distribution for muons coming from the primary vertex. The late component can be removed by applying a proper time window, that has to be optimized tower by tower. However the timing information are not used for the current studies, therefore better results are expected in the future.
\subsection{Tracks reconstruction performance}
\label{sec:track}
\begin{figure}[htbp]
\centering
    \includegraphics[width=0.48\textwidth]{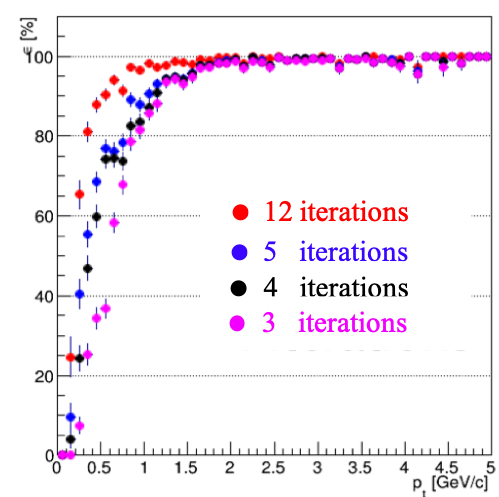}
    \includegraphics[width=0.48\textwidth]{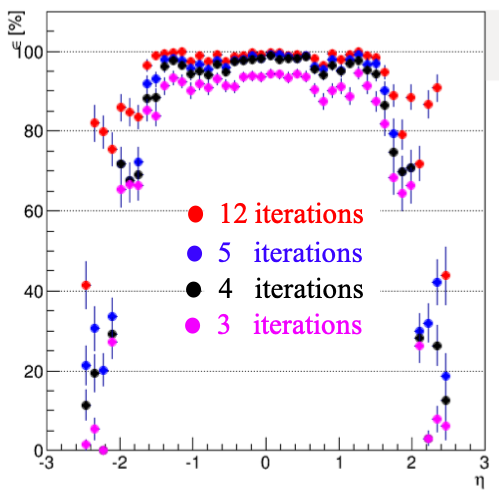}
    \caption{Tracking efficiency for different numbers of iterations as a function of the transverse momentum $p_T$ (left) and the psudorapidity $\eta$ (right) in a sample of muons.
    \label{fig:traking-iterations}}
\end{figure}

Hits surviving the selection discussed in Section~\ref{sec:dete} are used as an input to a parallel Kalman filter, as implemented in the ILCRoot framework. Pattern recognition and track finding are performed simultaneously with an iterative procedure with increasing search windows for hits in the following tracker layers at each iteration. 
This method guarantees high tracking efficiency, but requires significant computing resources and long processing times. With the available code, only four iterations were feasible for events with beam-induced background. Figure~\ref{fig:traking-iterations} shows the tracking efficiency evaluated in a sample of muons as a function of transverse momentum ($p_{\mathrm{T}}$) and pseudorapidity ($\eta$) for different numbers of iterations. The tracking efficiency is defined as the fraction of generated particles in the tracker geometrical acceptance that are matched to a good-quality reconstructed track. The effect of fewer iterations in the track finding procedure is a drop in efficiency for low-$p_T$ and high-$\eta$ particles. Four iterations represented a good trade-off between an acceptable reconstruction efficiency and a practical processing time. As an example, the tracking of one beam-induced background event, $\sim$60000 tracks with $p_T > 0.5$ GeV, takes approximately one hour on an Intel Xeon CPU E5-2665 2.40GHz with 32GB RAM in the case of four iterations with a CPU time that increases non-linearly with the number of iterations.

\begin{figure}[htbp]
\centering
    \includegraphics[width=0.48\textwidth]{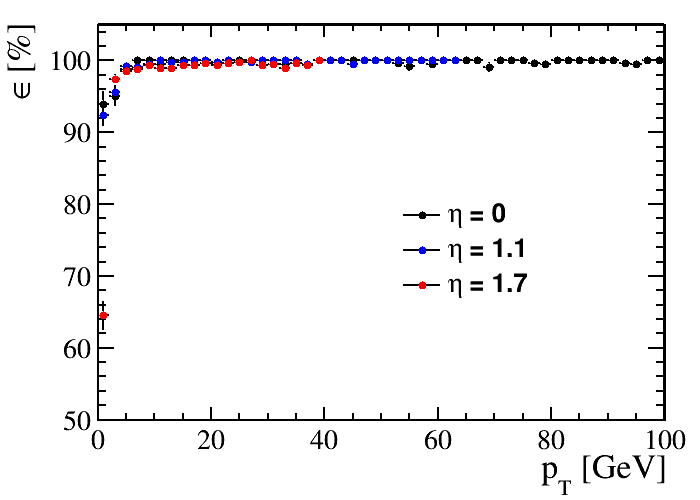}
    \includegraphics[width=0.48\textwidth]{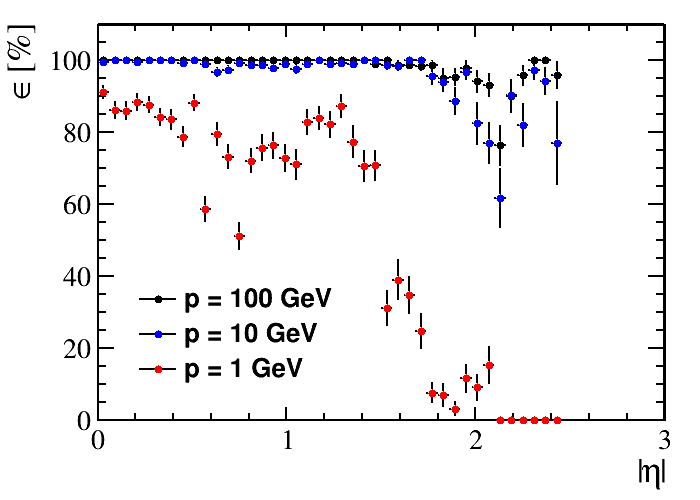}
    \caption{Muon tracking efficiency as a function of $p_{\mathrm{T}}$ for three representative $|\eta|$ values (left) and as a function of $|\eta|$ for three different momenta (right).}
    \label{fig:traking-eff}
\end{figure}

\begin{figure}[htbp]
\centering
    \includegraphics[width=0.48\textwidth]{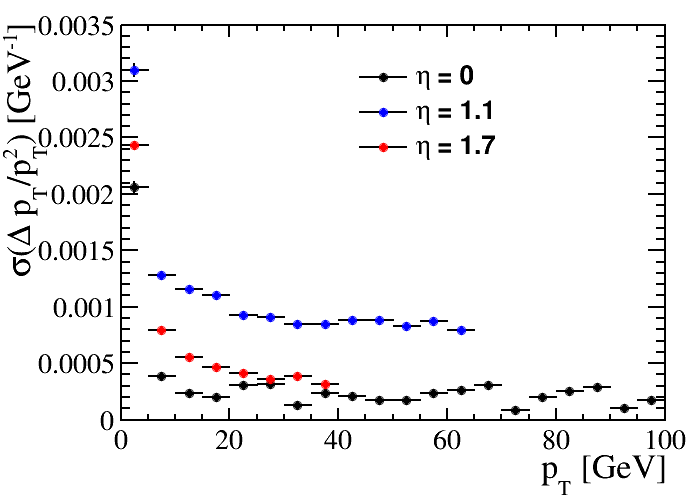}
    \includegraphics[width=0.48\textwidth]{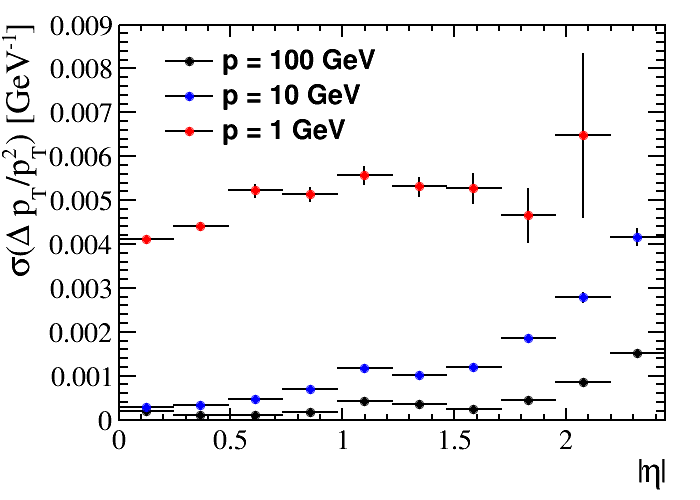}
    \caption{Track $p_T$ resolution as a function of $p_T$ (left) for three representative values of $|\eta|$ and as a function of $|\eta|$ (right) for three ranges of muon momenta. Each bin has been calculated as the width of the distribution $\Delta p_T/p_T^2$, where $\Delta p_T$ is the difference between the generated muon $p_T$ and the $p_T$ of the corresponding reconstructed track. 
    }
    \label{fig:traking-res}
\end{figure}

The track reconstruction performance is assessed in samples of single muons generated at different momenta and pseudorapidities. 
In the left panel of Figure~\ref{fig:traking-eff} is shown the tracking efficiency as a function of the transverse momentum for three representative values of $|\eta|$: the tracker central region, the forward region and the crack between the two. As expected, there is a loss of efficiency at low momentum, which is more evident on the right panel, where the efficiency is plotted as a function of $|\eta|$ for three different momentum values. Low efficiency at high $\eta$ is expected both due to the lower performance of the reconstruction algorithm and the limited coverage of the detector because of the shielding nozzles.
The relative transverse momentum resolution of the reconstructed tracks is reported in Figure~\ref{fig:traking-res} for samples of single muons. The left panel shows the track $p_{\mathrm{T}}$ resolution as a function of $p_{\mathrm{T}}$ in different $|\eta|$ regions. In the central and forward regions the $p_{\mathrm{T}}$ resolution is lower than $5\times 10^{-4}$ GeV$^{-1}$ for $p_{\mathrm{T}} > 10$ GeV, while tracks reconstructed in the crack between the central layers and the forward disks present a degraded resolution. The right panel shows the $p_{\mathrm{T}}$ resolution as a function of pseudorapidity for muons with $p =$ 1, 10, 100 GeV. The resolution degrades visibly for tracks of low momentum and high $|\eta|$.

\subsection{Jet reconstruction and identification performance}
\label{sec:jet}
Jet reconstruction was not part of the ILCRoot package and a simple dedicated algorithm that takes into account the high yield of particles coming from the beam-induced background has been developed. 
Jet reconstruction is performed using the calorimeter clusters, which are selected according to the following algorithm:
\begin{enumerate}
    \item the calorimeter detector is divided in several pseudorapidity regions of equal width;
    \item in each region the mean $\langle E \rangle$ and the standard deviation $\sigma_E$ of the calorimeter cluster energies are calculated;
    \item calorimeter clusters with an energy $E$ higher than $\langle E \rangle + 2 \cdot \sigma_E$ are selected;
    \item the energy of the selected clusters is corrected by subtracting the mean value $\langle E \rangle$ of the corresponding region.
\end{enumerate}
This selection algorithm is designed to remove, at least in part, the beam-induced background which is diffused in the calorimeter as shown in Figure ~\ref{fig:calo-occupancy}. In this way, signals originated by particles coming from the primary interaction, are identified as energy deposition significantly above the mean background energy.

Selected calorimeter clusters are then used as inputs to the jet clustering algorithm: a cone algorithm~\cite{jet-cone} with a radius parameter of $R=0.5$ is employed, where $R$ is the distance in the pseudorapidity ($\eta$) and azimuthal angle ($\phi$) plane between the jet axis and each cluster. Calorimeter clusters with an energy greater than 2 GeV are considered as seeds. The raw jet energy is obtained as the sum of the energies of the clusters belonging to the jet.

In order to evaluate the jet energy corrections needed to assess the jet algorithm performance, a simulated sample of $\mu^+ \mu^- \rightarrow H( \rightarrow b \bar{b}) \nu \bar{\nu}$ events at $\sqrt{s}=1.5$ TeV is used. This physics process has been generated at particle level with Pythia~8~\cite{pythia} and the full simulation with beam-induced background is employed to obtain the detector response. Truth-level jets are clustered by using truth-level stable particles as inputs to the cone algorithm. Then a reconstructed, detector-level jet is matched with a truth-level jet if the distance in the $(\eta,\phi)$ space between their axes is less than 0.5. Reconstructed jets that fulfill this requirement are indicated as truth-matched jets in the text.

\begin{figure}[htbp]
\centering
    \includegraphics[width=0.6\textwidth]{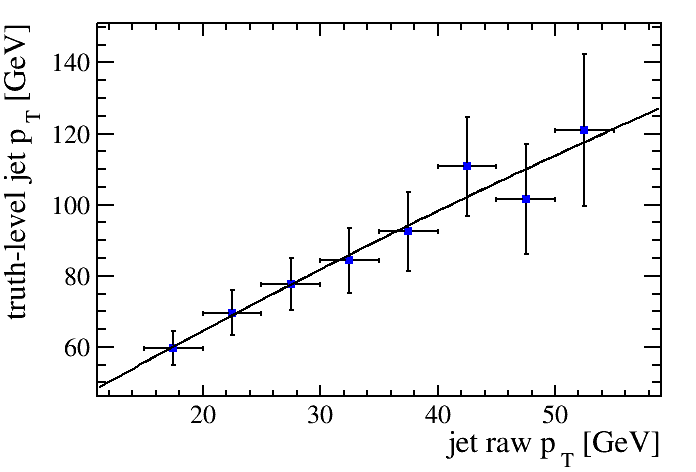}
    \caption{Truth-level jet $p_\mathrm{T}$ as a function of the raw $p_\mathrm{T}$ of reconstructed jets.}
    \label{fig:jet_scale}
\end{figure}

The truth-level jet transverse momentum ($p_{\mathrm{T}}$) is shown in Figure~\ref{fig:jet_scale} as a function of the average jet raw $p_{\mathrm{T}}$.
It is evident that the truth-level jet $p_\mathrm{T}$ is strongly correlated with the raw jet $p_\mathrm{T}$ demonstrating the correctness of the reconstruction.
A polynomial fit to the graph is performed in order to obtain the jet energy correction function, which is applied to the reconstructed-level jets to determine the nominal jet energy starting from the raw jet energy.

\begin{figure}[htbp]
\centering
    \includegraphics[width=0.48\textwidth]{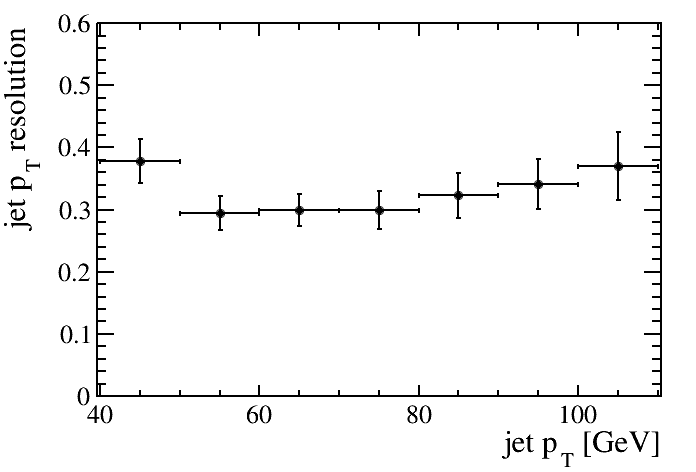}
    \includegraphics[width=0.48\textwidth]{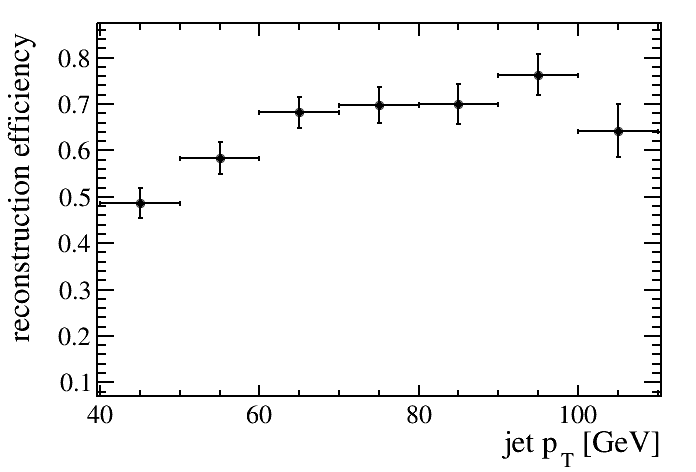}
    \caption{Left: average jet transverse momentum resolution as a function of the jet transverse momentum. Right: jet reconstruction efficiency as a function of the jet transverse momentum.}.
    \label{fig:jet_performance}
\end{figure}

The jet transverse momentum resolution $\frac{\Delta p_\mathrm{T}}{p_\mathrm{T}}$ is defined as the difference between the jet $p_\mathrm{T}$ and the matched truth-level jet $p_\mathrm{T}$, divided by the truth-level jet $p_\mathrm{T}$. The average $\frac{\Delta p_\mathrm{T}}{p_\mathrm{T}}$ as a function of the jet transverse momentum is shown in Figure~\ref{fig:jet_performance} (left) for simulated $b$-jets . The resolution varies from $30\%$ to $40\%$ depending on the jet $p_{\mathrm{T}}$.

The jet reconstruction efficiency, the number of reconstructed jets matched with truth-level jets divided by the total number of truth-level jets is shown, as a function of the jet transverse momentum, on the right of Figure~\ref{fig:jet_performance}.
The efficiency is around $50\%$ for low-$p_\mathrm{T}$ jets and around $70\%$ for $p_\mathrm{T}>60$ GeV.

The rate of fake jets is not really meaningful: the algorithm is not optimized and a better jet reconstruction will be achievable in the future by implementing a Particle Flow algorithm that considers both tracks and calorimeter clusters as inputs. 
In order to give an idea of the performance on the current reconstruction, the number of reconstructed jets not matched with a truth-level jet divided by the total number of reconstructed jets is evaluated. The fake-rate is found to be the order of $25\%$ for jets with $p_\mathrm{T}>40$ GeV. 

\section{Reconstruction of Higgs boson decay to $b$-jets}
The $H\to b \bar b$ final state is identified in two steps: the identification of $b-$jets and the selection of the $b \bar b$ resonance among physics background events.

\subsection{$b$-jet identification algorithm and performance}
 A $b$-jet tagging algorithm has been developed to reduce the background coming from light quark and gluon jets and from fake jets originated by several background sources in particular the beam-induced background. The algorithm is inspired by the one employed by the LHCb collaboration \cite{lhcb_tagging} and uses tracks inside the jet cone to identify decay vertices compatible with $b$-hadron decays. The algorithm has the following steps:
\begin{enumerate}
    \item for a given reconstructed jet, tracks inside the jet cone are selected, requiring for each track a $p_{\mathrm{T}}$ greater than 500 GeV and an impact parameter with respect to the $\mu^+\mu^-$ interaction point greater than 0.04 cm. Only good quality tracks are selected by requiring a minimum number of hits;
    \item 2-track vertices are formed by imposing the distance of closest approach between the tracks to be less than 0.02 cm. The  total $p_{\mathrm{T}}$ of the two tracks must be greater than 2 GeV;
    \item 3-track vertices are formed by linking two 2-track vertices that share one track;
    \item if at least one 3-tracks vertex is found, then the jet is tagged as a $b$-jet.
\end{enumerate}
This algorithm is very simple and cut-based. Presently higher performance methods are available which exploit machine learning techniques. In the future the most suitable algorithm, tuned for the process under study will be applied.

\begin{figure}[htbp]
\centering
    \includegraphics[width=0.48\textwidth]{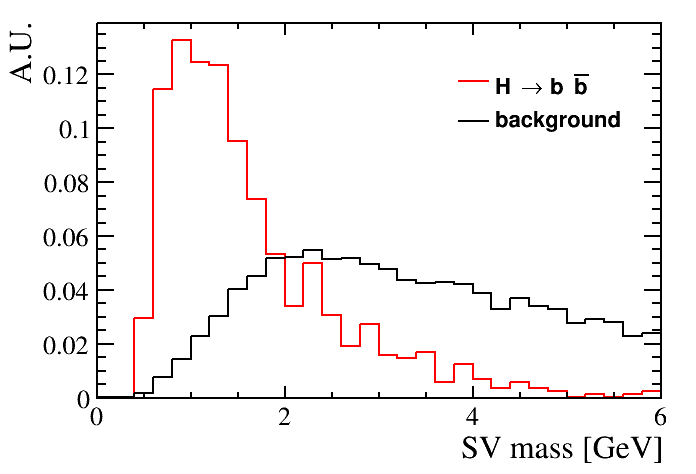}
    \includegraphics[width=0.48\textwidth]{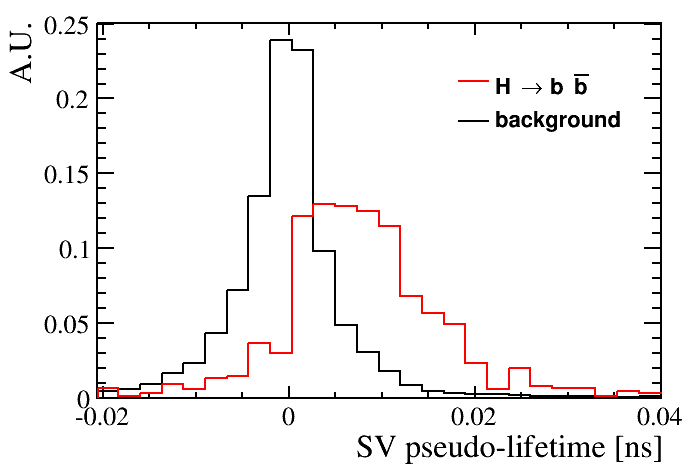}
    \caption{Left: invariant mass of tracks associated to SV for truth-matched $b$-jets from Higgs decay events (red) and for beam-induced background events (green). Right: pseudo-lifetime for truth-matched $b$-jets from Higgs decay events (red) and for beam-induced background (green). A negative pseudo-lifetime is assigned if $d$ projection with respect to the jet axis is negative.}
    \label{fig:b_tag}
\end{figure}

The limitations on the tracks reconstruction explained in~\ref{sec:track}, prevent application of the $b-$jet tagging algorithm to the full event, $\mu^+ \mu^- \rightarrow H( \rightarrow b \bar{b}) \nu \bar{\nu}$ with beam-induced background overlay. 
The overall performance is evaluated separately and then applied to the Higgs boson reconstruction.
In Figure \ref{fig:b_tag} the two most discriminating observables related to secondary vertex (SV) in truth-matched $b$-jets are compared with those related to SVs found in background events, in this particular case the beam-induced background. The two observables are the invariant mass of the tracks belonging to the SV and the pseudo-lifetime, defined as $\tilde{t} = \frac{m_{B^0} \cdot d}{p}$, where $m_{B^0}$ is the nominal $B^0$ mass, $d$ is the secondary vertex distance from the interaction point and $p$ is the overall momentum of the tracks associated to the SV. It is evident that it will be well within the performance of any algorithm to efficiently select $b-$jets.

\begin{figure}[htbp]
\centering
    \includegraphics[width=0.6\textwidth]{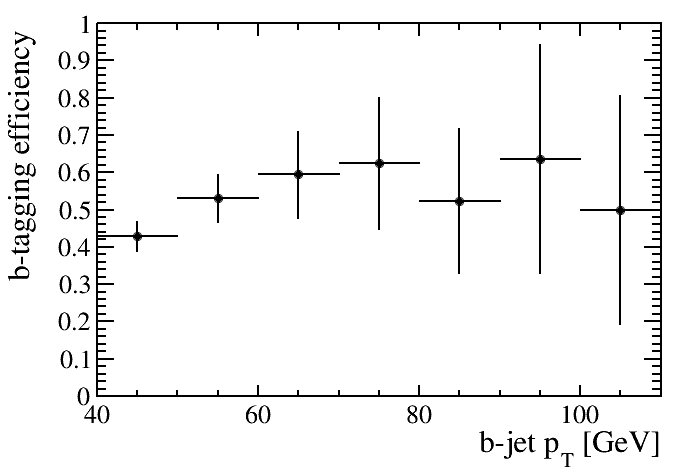}
    \caption{$b$-tagging efficiency as a function of the jet $p_{\mathrm{T}}$.}
    \label{fig:tag_eff}
\end{figure}

The $b$-tagging efficiency is defined as the number of tagged and reconstructed $b$-jets divided by the total number of reconstructed $b$-jets. The $b$-tagging efficiency as a function of the jet $p_{\mathrm{T}}$ is presented in Figure~\ref{fig:tag_eff}. 
It can be seen that the it varies from about $40\%$ to about $60\%$ at high jet $p_{\mathrm{T}}$. These values are very close to what is currently obtained in analyses of the LHC experiments. 
The misidentification rate is evaluated  on a statistical basis since it is not possible to determine it directly by using beam-induced background events due to tracking limitations.
The ($\eta,\phi$) space is divided in cone regions with $R=0.5$, in each cone SVs are reconstructed with the $b$-tagging algorithm. The ratio between the number of cones with at least one fake SV and the total number of cones is taken as the misidentification rate. It varies between $1\%$ and $3\%$ depending on the requirements on the tracks.
In this way the effect of the beam induced background on the Higgs boson identification is included through the b-tagging efficiency.
\subsection{$H \rightarrow b \bar{b}$ selection}
\begin{figure}[htbp]
\centering
    \includegraphics[width=0.6\textwidth]{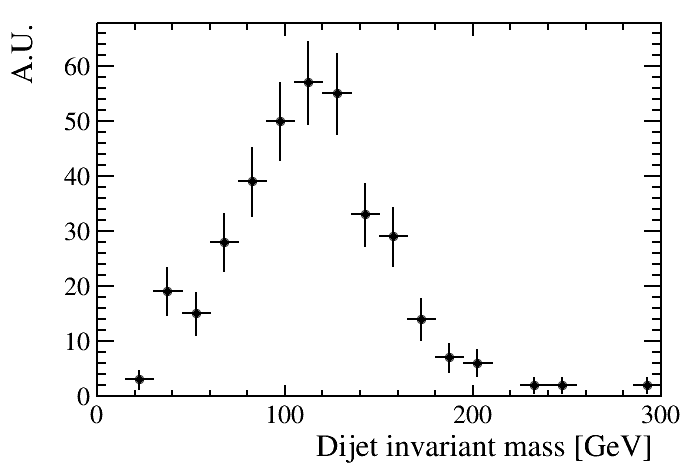}
    \caption{The di-jet invariant mass of truth-matched $b$-jets in $H \rightarrow b \bar{b}$ full-simulated events at $\sqrt{s}=1.5$ TeV.}
    \label{fig:higgs_mass}
\end{figure}

The $\mu^+ \mu^- \rightarrow H( \rightarrow b \bar{b}) \nu \bar{\nu}$ events generated at $\sqrt{s}=1.5$ TeV are reconstructed by using a simple procedure. 
The fiducial region is defined by two jets with $p_{\mathrm{T}}>40$ GeV and pseudorapidity in the range $-2.5 < \eta < 2.5$.
Therefore, the first requirement is to have two reconstructed and $b$-tagged jets in the fiducial region.
The acceptance, $A$, is defined as the number of events with two $b$-jets from Higgs boson decay in the fiducial region divided by the number of generated $H \rightarrow b \bar{b}$ events. At $\sqrt{s}=1.5$ TeV the acceptance is found to be $A=0.35$. The selection efficiency ($\epsilon$) is defined as the number of fully reconstructed $H \rightarrow b \bar{b}$ events divided by the number of events with two truth-level $b$-jets from Higgs boson decay in the fiducial region, and it is found to be $\epsilon = 15\%$.
The reconstructed di-jet invariant mass of truth-matched $b$-jets in $H \rightarrow b \bar{b}$ events is shown in Figure~\ref{fig:higgs_mass}. The presence of the beam-induced background could have jeopardized this distribution making the Higgs boson identification very difficult. Instead the Higgs boson peak is clearly visible in the di-jet invariant mass.
\section{Higgs Boson coupling to $b-$quark}
In addition to demonstrating that one of the most critical decay channels, $H\to b \bar b$, is reconstructed efficiently at a muon collider, the achievable sensitivity on the couplings is evaluated for three different center-of-mass energies.
\subsection{Higgs Boson coupling to $b-$quark at $\sqrt{s}=1.5$ TeV}
\label{sec:Hcoupling}
In this section the sensitivity of the measurement of the Higgs boson coupling to $b$ quarks ($g_{Hbb}$) in $\mu\mu$ collisions at $\sqrt{s}=1.5$ TeV is studied. 
Since the Higgs boson production at $\sqrt{s}=1.5$ TeV is dominated by $WW$ fusion \cite{mc-pre} the $H \rightarrow b \bar{b}$ production cross section ($\sigma$) is related to the Higgs boson couplings as follows:
\begin{equation}
    \label{eq:coupling}
    \sigma = \sigma(\nu \nu H) \cdot BR(H \rightarrow b \bar{b}) = \frac{g^2_{HWW} g^2_{Hbb}}{\Gamma_{H}},
\end{equation}
where $g_{HWW}$ is the coupling of the Higgs boson to the $W$ boson and $\Gamma_{H}$ is the Higgs boson width.
The uncertainty on $g^2_{Hbb}$ is, therefore, related to the measured cross section.

The number of observed $H \rightarrow b \bar{b}$ events ($N$) is given by:
\begin{equation}
    N = A \cdot \epsilon \cdot \sigma \cdot \mathcal{L} \cdot t,
\end{equation}
where $A$ is the acceptance, $\epsilon$ is the selection efficiency, $\mathcal{L}$ is the instantaneous luminosity and $t$ is the data-taking time.
By using a sample of events generated with Pythia~8, which are fully simulated and reconstructed following the procedure described above, the $H \rightarrow b \bar{b}$ cross section at $\sqrt{s}=1.5$ TeV is found to be $\sigma=203$ fb. Therefore, with an instantaneous luminosity of $\mathcal{L} = 1.25 \cdot 10^{34} $ cm$^{-2}$s$^{-1}$ in a data taking period of four Snowmass years ($t=4 \cdot 10^7$ s) the number of observed Higgs boson events decaying to $b \bar b$ is $N=5500$.

The uncertainty on the cross section determination is calculated by assuming $A$, $\epsilon$, $\mathcal{L}$, $t$ known with a good precision, and it is given, approximately,  by the number of events after all the selections:
\begin{equation}
   \label{eq:delta_xs}
    \frac{\Delta \sigma}{\sigma} \simeq \frac{\sqrt{N+B}}{N},
\end{equation}
where $B$ is the expected number of background events. In principle, with a combined fit to the di-jet invariant mass and discriminating kinematic variables, it should be possible to disentangle the signal from the background, thus reducing the uncertainty on the cross section. In this study, a conservative estimate based on equation \ref{eq:delta_xs} is used to evaluate the uncertainty on the cross section.

The contribution to the background events comes from two sources: physics processes and uncorrelated background. The first contribution is evaluated with Monte Carlo simulation.  The Standard Model processes considered as possible background are listed in Table~\ref{tab:bkg}.
\begin{table}[htbp]
    \centering
    \begin{tabular}{c}
         \hline
         Process          \\ 
         \hline
         $\mu^+\mu^-\to \gamma^*/Z\to q \bar q$    \\
         $\mu^+\mu^-\to \gamma^*/Z \gamma^*/Z \to q \bar q$ +X  \\
         $\mu^+\mu^-\to \gamma^*/Z \gamma \to q \bar q \gamma$  \\
        \hline
    \end{tabular}
    \caption{List of Standard Model processes considered in the background estimation for the $H \rightarrow b \bar{b}$ cross section measurement. $q$ indicates a generic quark.}
    \label{tab:bkg}
\end{table}
The number of background events with $b\bar{b}$ final states have been estimated by applying reconstruction and $b$-tagging efficiencies to generator-level samples produced with Pythia~8.
The contribution of light quark events, for example those with two $b-$tags in the final state, due mainly to beam-induced background, has been calculated applying the same procedure used for physics processes. In this case, however, the tagging mis-identification rate is used instead of the $b$-tagging efficiency. The fake-jet contribution, after the $b$-tagging requirements, is considered negligible. 
The number of expected background events after considering all the contributions, is $B=6700$. 
At this point it is possible to evaluate the uncertainty on the cross section by using equation~\ref{eq:delta_xs}, $\frac{\Delta \sigma}{\sigma}=2.0 \%$.

According to equation~\ref{eq:coupling}, it is necessary to also know $g_{HWW}$ and $\Gamma_{H}$ in order to determine $g_{Hbb}$. Knowledge of the $H \rightarrow b \bar{b}$ cross section is not enough. In the following, it is assumed that $g_{HWW}$ and $\Gamma_{H}$ can be measured with the same level of precision expected by CLIC at $\sqrt{s}=1.4$ TeV \cite{higgs_clic}. This is justified by the fact that the selection of muonic final states at a muon collider is analogous to that at electron-positron accelerators, since the beam-induced background stops at the calorimeters and is not expected in muon detectors. 
Therefore the uncertainty on the coupling can be obtained with:
\begin{equation}
    \frac{\Delta g_{Hbb}}{g_{Hbb}} = \frac{1}{2} \sqrt{ \left(\frac{\Delta \sigma}{\sigma} \right)^2 + \left(\frac{\Delta \frac{g^2_{HWW}}{\Gamma_H}}{\frac{g^2_{HWW}}{\Gamma_H}} \right)^2 },
\end{equation}
where the uncertainty on $\frac{g^2_{HWW}}{\Gamma_H}$ has been extracted from the CLIC study \cite{higgs_clic} and scaled for the lower integrated luminosity assumed for the muon collider at $\sqrt{s}=1.5$ TeV.
The expected sensitivity on the Higgs coupling to $b$ quark at $\sqrt{s}=1.5$ TeV is then found to be $\frac{\Delta g_{Hbb}}{g_{Hbb}}=1.9\%$.

\subsection{Higgs Boson coupling to $b$ quarks at
$\sqrt{s}=3$ TeV and $\sqrt{s}=10$ TeV}
The procedure used in~\ref{sec:Hcoupling} is also applied to evaluate the sensitivity to the $g_{Hbb}$ coupling when it is measured in muon collisions at $\sqrt{s}=3.0$ TeV and $\sqrt{s}=10$ TeV.
This approach is very conservative. The efficiencies obtained with the full simulation at $\sqrt{s}=1.5$ TeV are used for the higher center-of-mass energy cases, with the proper scaling to take into account the different kinematic region. At higher $\sqrt{s}$ the tracking and the calorimeter detectors are expected to perform significantly better since the yield of the beam-induced background decreases with $\sqrt{s}$ as demonstrated in~\cite{mc-pre}. 
The uncertainty on $\frac{g^2_{HWW}}{\Gamma_H}$ at $\sqrt{s}=3.0$ TeV is taken from the CLIC study at the same center-of-mass energy \cite{higgs_clic}. At $\sqrt{s}=10$ TeV this uncertainty is assumed equal to the one at $\sqrt{s}=3.0$ TeV, for the moment this is the only estimated number and, following the conservative approach that drives this work, it is used as it is. It is reasonable to imagine that when the full Higgs boson couplings analysis will be done at $\sqrt{s}=10$ TeV it will improve.

The instantaneous luminosity, $\mathcal{L}$, at different $\sqrt{s}$ are taken from~\cite{map-lumi}. The integrated luminosity $\mathcal{L}_{int}$, is calculated by using the standard four Snowmass years. The acceptances, $A$, the number of signal events, $N$, and background, $B$, are determined with simulation. The uncertainties on $\sigma$ and $g_{Hbb}$ are calculated and summarized in Table~\ref{tab:err_xs} with all relevant inputs. The resulting relative uncertainty on the coupling is $1.0\%$ at $\sqrt{s}=3.0$ TeV and $0.91\%$ at $\sqrt{s}=10$ TeV.
It has to be noted that the result at $\sqrt{s}=10$ TeV is dominated by the error on $\frac{g^2_{HWW}}{\Gamma_H}$, which is assumed equal to the one used at $\sqrt{s}=3$ TeV.
\begin{table}[htbp]
    \centering
    \begin{tabular}{c|c|c|c|c|c|c|c|c|c}
         \hline
         $\sqrt{s}$ & $A$ & $\epsilon$ & $\mathcal{L}$ &  $\mathcal{L}_{int}$ & $\sigma$ & $N$ & $B$ & $\frac{\Delta \sigma}{\sigma}$ & $\frac{\Delta g_{Hbb}}{g_{Hbb}}$ \\ 
         
         [TeV] & [\%] & [\%] & [cm$^{-2}$s$^{-1}$] & [ab$^{-1}$] & [fb] &  &  & [\%] & [\%] \\
         \hline
         1.5 & 35 & 15 & $1.25 \cdot 10^{34}$ & 0.5 & 203 & 5500  & 6700 & 2.0 & 1.9 \\
         
         3.0 & 37 & 15 & $4.4 \cdot 10^{34}$ & 1.3 & 324 & 33000 & 7700 & 0.60 & 1.0 \\
         
         10 & 39 & 16 & $2 \cdot 10^{35}$ & 8.0 & 549 &  270000 & 4400 & 0.20 & 0.91 \\
        \hline
    \end{tabular}
    \caption{\label{tab:err_xs} Summary of the parameters used as inputs for the determination of the Higgs coupling to $b$ quarks. The data taking time is assumed of $4 \cdot 10^{7}$ s. The parameter definitions are given in the text.}
\end{table}
\section{Comparison to other future machines}
The direct comparison of the results obtained on $\frac{\Delta g_{Hbb}}{g_{Hbb}}$ at muon collider with other colliders, as done in~\cite{Hcouplings}, is not available yet. In order to have an idea of the potential of an experiment at a muon collider, these results are compared to those published by CLIC~\cite{higgs_clic}. CLIC numbers are obtained with a model-independent multi-parameters fit. In addition, the fit is performed in three stages, taking the  statistical uncertainties obtainable at the three considered energy successively into account. This means that each new stage includes all measurements of the previous stages and is represented in Table~\ref{tab:comparison} with a "+" in the integrated luminosity. 

The muon collider results are not complete, since not all the necessary parameters are determined. They are based on assumptions that are very conservative, as discussed in the previous sections. Data sample at the three center-of-mass energies are treated as independent, and not taken successively into account. This means that at $\sqrt{s}=3$~TeV the precision achieved by the experiment at muon collider uses 4 data-taking years while the CLIC number includes also the 4 years at $\sqrt{s}=350$~GeV.
\begin{table}[htbp]
    \centering
    \begin{tabular}{l|c|c|c}
         \hline
           & $\sqrt{s}$ [TeV] & $\mathcal{L}_{int}$ [ab$^{-1}$]& $\frac{\Delta g_{Hbb}}{g_{Hbb}}$ [\%]\\  \hline
           \multirow{3}{*}{Muon Collider}
           &1.5 & 0.5 & 1.9 \\
           &3.0 & 1.3 & 1.0 \\
           &10  & 8.0 & 0.91 \\
  \hline
  \multirow{3}{*}{CLIC}
           &0.35& 0.5 & 3.0 \\
           &1.4 & +1.5 & 1.0 \\
           &3.0 & +2.0 & 0.9 \\
     \hline       
    \end{tabular}
    \caption{\label{tab:comparison} Relative precision on Higgs boson coupling to $b-$quark at muon collider and at CLIC. The difference on how the numbers are obtained by the two experiments is described in the text.}
\end{table}    
\section{Summary and Conclusion}
A detailed study of the Higgs boson decay to $b-$jets at $\sqrt{s}=1.5$~TeV is presented, based on full simulation of the physics process and the beam-induced background. Physics performance of the tracking and calorimeter detectors are discussed together with new ideas to mitigate the effect of the beam-induced background. The Higgs boson decay to $b-$jets is efficiently reconstructed demonstrating that the beam-induced background does not jeopardize physics performance of an experiment at muon collider.
The uncertainty on the Higgs boson coupling to $b-$quark is determined under several assumptions and compared to the results obtained by CLIC in similar conditions showing the potential of the muon collider. It has to be noted that CLIC has quoted the best precision on $g_{Hbb}$~\cite{Hcouplings}. 
\acknowledgments
We owe a huge debt of gratitude to Anna Mazzacane and Vito Di~Benedetto (FNAL) who made this study possible by providing us with the MAP's software framework and simulation code. We acknowledge support from Istituto Nazionale di Fisica Nucleare and from Department of Physics and Astronomy of the University of Padova. We thank the VenetoCloud for providing computing resources and Lucio Strizzolo and the IT team of INFN Trieste for their technical support with the local cloud computing.

\end{document}